\documentclass[aps,prl,showpacs,twocolumn,groupedaddress,floatfix]{revtex4}

\usepackage{graphicx}
\usepackage{color}
\def\beq{\begin{eqnarray}}
\def\eeq{\end{eqnarray}}
\def\to{\widetilde{\Omega}}

\begin{document}

\title{A new method for the solution of the Schr\"odinger equation}
\author{Paolo Amore}
\email{paolo@ucol.mx}
\author{Alfredo Aranda}
\email{fefo@cgic.ucol.mx}
\affiliation{Facultad de Ciencias, Universidad de Colima, \\
Bernal D\'{\i}az del Castillo 340, Colima, Colima, M\'exico} 
\author{Arturo De Pace}
\email{depace@to.infn.it}
\affiliation{Istituto Nazionale di Fisica Nucleare, sezione di Torino, \\
via P. Giuria 1, I-10125, Torino, Italy}

\begin{abstract}    
We present a new method for the solution of the Schr\"odinger 
equation applicable to problems of non-perturbative nature. The method works by
identifying three different scales in the problem, which then are treated
independently: An asymptotic scale, which depends uniquely on the form of the
potential at large distances; an intermediate scale, still characterized by an
exponential decay of the wave function and, finally, a short distance scale, 
in which the wave function is sizable. The key feature of our method is the
introduction of an arbitrary parameter in the last two scales, which is then
used to optimize a perturbative expansion in a suitable parameter. We apply the
method to the quantum anharmonic oscillator and find excellent results. 
\end{abstract}
\pacs{03.65.Ge,02.30.Mv,11.15.Bt,11.15.Tk}      
\maketitle


In this Letter we present a new method to find approximate solutions to
eigenvalue problems in Quantum Mechanics. 
In particular, this method is applicable to problems where ordinary 
perturbation theory generates divergent asymptotic series. 
Actually, improving the convergence of the standard Rayleigh-Schr\"odinger
perturbative  expansion has been the subject of many studies in the past (see,
e.~g., Refs.~\cite{Hatsuda:1996vp,BB96,Gui95,Jan95} and references therein) and
many variants of ``optimized expansions'' have been proposed.

Among them we would like to single out the Linear Delta Expansion (LDE) 
\cite{lde}, which is at the base of our developments.
The LDE has been extensively applied in many different settings with varying 
degrees of success. For example, in Ref.~\cite{blencowe} it has been used to 
analyze disordered systems. In Ref.~\cite{Jones:2000au} it has been applied to 
study the slow roll potential in inflationary models.
Pinto and collaborators have applied it to the 
Bose-Einstein condensation problem~\cite{Kneur:2002dn}, the
$O(N)(\phi^2)^2_{3d}$ model~\cite{Kneur:2002kq}, the Walecka 
model~\cite{Krein:1995rp} and to the $\phi^4$ theory at high
temperature~\cite{Pinto:1999py}.  More recently the LDE has been applied with
success to the study of classical nonlinear systems \cite{AA1:03,AA2:03}. 
Detailed references on the method can be found in these works.

Here, we illustrate the method we propose by applying it to the quantum
anharmonic oscillator, which is the usual benchmark to test any
non-perturbative method. 

The cornerstone of the method is the identification of three different scales 
in the problem, which give rise to different behaviors of the wave function.
Keeping in mind the standard separation of the Hamiltonian in a unperturbed
piece (the harmonic oscillator) and a perturbative one (the anharmonic term),
one can recognize that at very large distances the wave function assumes its 
asymptotic behavior: This is completely determined by the anharmonic potential
and it is the same for all (ground and excited) states; 
at intermediate distances the wave function still decays exponentially, but now
governed by the harmonic term; finally, there is a short distance scale, 
in which the wave function is sizable. 

The key feature of our method is the introduction of an arbitrary parameter in 
the last two scales, which is then used, in the LDE spirit, to 
optimize a perturbative expansion in a suitable parameter. 

Consider the  Schr\"odinger equation
\beq
  \left[ - \frac{\hbar^2}{2m} \frac{\partial^2}{\partial x^2} +  \frac{m
    \omega^2}{2} x^2 + \frac{\mu}{4} x^4 \right] \psi_n(x) = E_n \psi_n(x),
\label{eq2}
\eeq
where $\mu$ is the anharmonic coupling.
The asymptotic behavior of $\psi_n(x)$ in the region of large $x$ is
determined by substituting the ansatz $\psi_n(x) \propto e^{- \gamma  |x|^p}$
into Eq.~(\ref{eq2}). One obtains $p=3$ and 
$\gamma = \sqrt{\mu m/2}/3\hbar$.  

In order to make the three scales explicit in the wave function, we write
\beq
  \psi_n(x) =   e^{- \gamma |x|^3 - \beta x^2 } \xi_n(x),
\label{eq3}
\eeq
where the exponential takes care of the correct behavior in the limit 
$|x| \rightarrow \infty$.  
Notice that the quadratic term in the exponential does not affect the behavior
at large distances, but dominates at scales where $|x| < \beta/\gamma$. 
Here $\beta = m \sqrt{\omega^2+\Omega^2}/2\hbar$ is the coefficient of a 
harmonic oscillator of frequency
$\tilde{\Omega}\equiv\sqrt{\omega^2+\Omega^2}$, where $\Omega$ is an arbitrary 
parameter introduced by hand; $\xi_n$ is a well-behaved function, which
fulfills the equation \footnote{We are considering only the region $x>0$.  
The other region will be obtained by using the symmetry properties of the wave
function.}: 
\beq
  \xi_n''(x) &-& \left[ \frac{\sqrt{2 m \mu}}{\hbar} x^2 + 
    \frac{2m\to x}{\hbar} \right] \xi'_n(x) \nonumber \\
  &+& \left[ \frac{\sqrt{ 2\mu m^3 \to^2}}{\hbar^2} x^3 + 
    \frac{m^2 \Omega^2}{\hbar^2} x^2 - \frac{\sqrt{2m\mu}}{\hbar} x \right. 
    \nonumber \\ 
  &+& \left. \frac{2 m E_n}{\hbar^2} - \frac{m \to}{\hbar} \right] \xi_n(x)=0.
\label{eq4}
\eeq
Equation~(\ref{eq3}) has been introduced in order to single out three different regimes in the wave function: The purely asymptotic regime 
($|x| \rightarrow \infty$), where the cubic term in the exponential dominates;
the intermediate regime, where $|x|$ is not yet asymptotic but large enough to
expect the wave function to be exponentially damped; the regime of small $|x|$
where the physics is all contained in the $\xi$'s. The last two regimes will
display a dependence upon the arbitrary frequency $\Omega$, although in a quite 
different fashion (in fact, the intermediate regime displays a truly
non-perturbative dependence upon $\Omega$). 
In the limits $(\mu,\Omega) \rightarrow 0$ one obtains the equation
for the harmonic oscillator of frequency $\omega$, which admits polynomial
solutions (the Hermite polynomials). 

It is worth stressing that the energy $E_n$ in Eq.~(\ref{eq4}) is still the
true energy, since no approximation has been used to derive this  
equation.  

We observe in Eq.~(\ref{eq4}) that both the wave function $\xi_n(x)$ and the
energy $E_n$ depend in some nontrivial way upon the anharmonic coefficient
$\mu$. On  the other hand, the dependence of $\psi_n(x)$ and $E_n$ upon the
arbitrary frequency $\Omega$ is only fictitious, since this parameter does not
appear in the original equation~(\ref{eq2}). Nonetheless, we will show that
$\Omega$ can be used to generate an efficient expansion for the solution of 
Eq.~(\ref{eq2}).

Indeed we rewrite Eq.~(\ref{eq4}) as
\beq
  \xi_n''(x) &-& \left[ \frac{2 m \to x}{\hbar} \right] \xi'_n(x) + 
    \left[\frac{2 m E_n}{\hbar^2} - \frac{m \to }{\hbar} \right] \xi_n(x) 
    \nonumber \\ 
  &=& \delta \left\{ \frac{\sqrt{2 m \mu}}{\hbar} x^2 \xi'_n(x) 
    - \left[ \frac{\sqrt{ 2 \mu m^3 \to^2}}{\hbar^2} x^3 \right. \right. 
    \nonumber \\
  &+& \left. \left. \frac{m^2 \Omega^2}{\hbar^2} x^2 - 
    \frac{\sqrt{2 m \mu}}{\hbar} x \right] \xi_n(x) \right\} ,
\label{eq6}
\eeq
where the left hand side of Eq.~(\ref{eq6}) corresponds to the equation for a
harmonic oscillator of frequency $\to$. Following the spirit of the LDE, we
have introduced a parameter $\delta$, which is going to be used as a
power-counting device: When $\delta =1$, Eq.~(\ref{eq6}) reduces exactly to 
Eq.~(\ref{eq4}).

Although $\delta$ is not a small parameter we will treat the right hand side of 
Eq.~(\ref{eq6}) as a perturbation, writing down the following expansions:
\beq
  \xi_n(x) = \sum_{j=0}^\infty \delta^j \xi_{nj}(x), \ \  
    E_n = \sum_{j=0}^\infty \delta^j E_{nj}.
\label{eq6a}
\eeq
Combining Eqs.~(\ref{eq6}) and (\ref{eq6a}), one can generate a hierarchy of 
equations, corresponding to the different orders in $\delta$. 

Since we are doing perturbation theory, all the results, to any finite order in
the expansion, will depend upon the arbitrary frequency $\Omega$. 
Such dependence will therefore be minimized by applying the Principle of
Minimal Sensitivity (PMS) \cite{Ste81}, i.e. by requiring that a given
observable $O$ (the energy, for example)  be locally independent of $\Omega$: 
\beq
  \frac{\partial O}{\partial \Omega} = 0 .
\eeq
We will illustrate the method by explicitly showing the first two orders,
although we have obtained results up to the eighth order. 
To lowest order Eq.~(\ref{eq6}) reduces to the equation of a harmonic
oscillator of frequency $\to$, whose solutions are the Hermite polynomials,
\beq
  \xi_n(x) &=& H_n\left( \sqrt{ \frac{m \tilde{\Omega}}{\hbar}} x \right),
\label{eq:ord0a}
\eeq
while the energy eigenvalues are given by
\beq
  E_{n0} &=&  \hbar \tilde{\Omega} \left( n + \frac{1}{2} \right) ; \ \ 
    n = 0,1,\dots .
\label{eq:ord0b}
\eeq
To first order we have the equation:
\beq
  \xi_{n1}''(x) &-& \left[ \frac{2 m \tilde{\Omega} x}{\hbar} \right] 
    \xi'_{n1}(x) + \frac{2 m \tilde{\Omega}}{\hbar} n \xi_{n1}(x) \nonumber \\
  &=& \left\{ - \frac{2 m E_{n1}}{\hbar^2} + \frac{\sqrt{2m \mu}}{\hbar} x 
    \right. \nonumber \\
  &-& \left. \frac{m^2 \Omega^2}{\hbar^2} x^2 - \frac{\sqrt{2 \mu m^3
    \to^2}}{\hbar^2} x^3 \right\} \xi_{n0}(x) .
\label{eq:ord1}
\eeq
Although such equation is valid for any state of the AHO, for
illustrative purposes we will now consider only the ground state, for
which $\xi_{00}(x) = 1$.
Then, for the case $n=0$, the solution of Eq.~(\ref{eq:ord1}) is a polynomial of  
order $3$ and can be cast in terms of unknown coefficients as:
\beq
  \xi_{01}(x) &=&  a_0 + a_1 x + a_2 x^2 + a_3 x^3 .
\label{eq:ord1a}
\eeq
The coefficient $a_0$ is not determined by Eq.~(\ref{eq:ord1}) and we impose 
$a_0 = 0$, since it corresponds to the same functional form of $\xi_{00}$ 
\footnote{This is somewhat analogous of the procedure employed in the 
Lindstedt-Poincar\'e method to get rid of ``secular terms'' in the solutions 
\protect\cite{AA1:03,AA2:03}.}.
By substituting this polynomial into Eq.~(\ref{eq:ord1}) one gets the
coefficients 
\beq
  a_1 = 0 ; \ \ a_2 &=& \frac{m \Omega^2}{4 \hbar \tilde{\Omega}} ; \ \ 
    a_3 = \frac{1}{3 \hbar} \sqrt{\frac{m \mu}{2}} 
\label{eq:ord1b}
\eeq
and the energy
\beq
  E_{01} &=& - \frac{\hbar \Omega^2}{4 \tilde{\Omega}} .
\label{eq:ord1c}
\eeq
Therefore, up to first order, we get the energy
\beq
  E_{0}^{(1)} &=&  \frac{\hbar \tilde{\Omega}}{2} - 
    \frac{\hbar \Omega^2}{4 \tilde{\Omega}} .
\label{eq:ord1e}
\eeq
It is important to notice that the wave function obtained up to first order
does not have nodes, a desirable result for the wave function of the ground
state. 

The PMS to this order yields the solution $\Omega=0$ and the corresponding
energy 
\beq
  \left. E_{0}^{(1)}\right|_{\text{PMS}} &=& \frac{\hbar \omega}{2} .
\eeq

Here we would like to stress a point: We could have obtained a solution similar
to that of Eq.~(\ref{eq:ord1a}) by direct application of the 
Rayleigh-Schr\"odinger perturbation theory to the wave functions of the harmonic 
oscillator. However, had we applied the LDE directly to the
Rayleigh-Schr\"odinger expansion, we would not have been able to reproduce the
correct asymptotic behavior of the wave function \cite{Hatsuda:1996vp}.  

\begin{figure}[t]
\includegraphics[width=8.5cm]{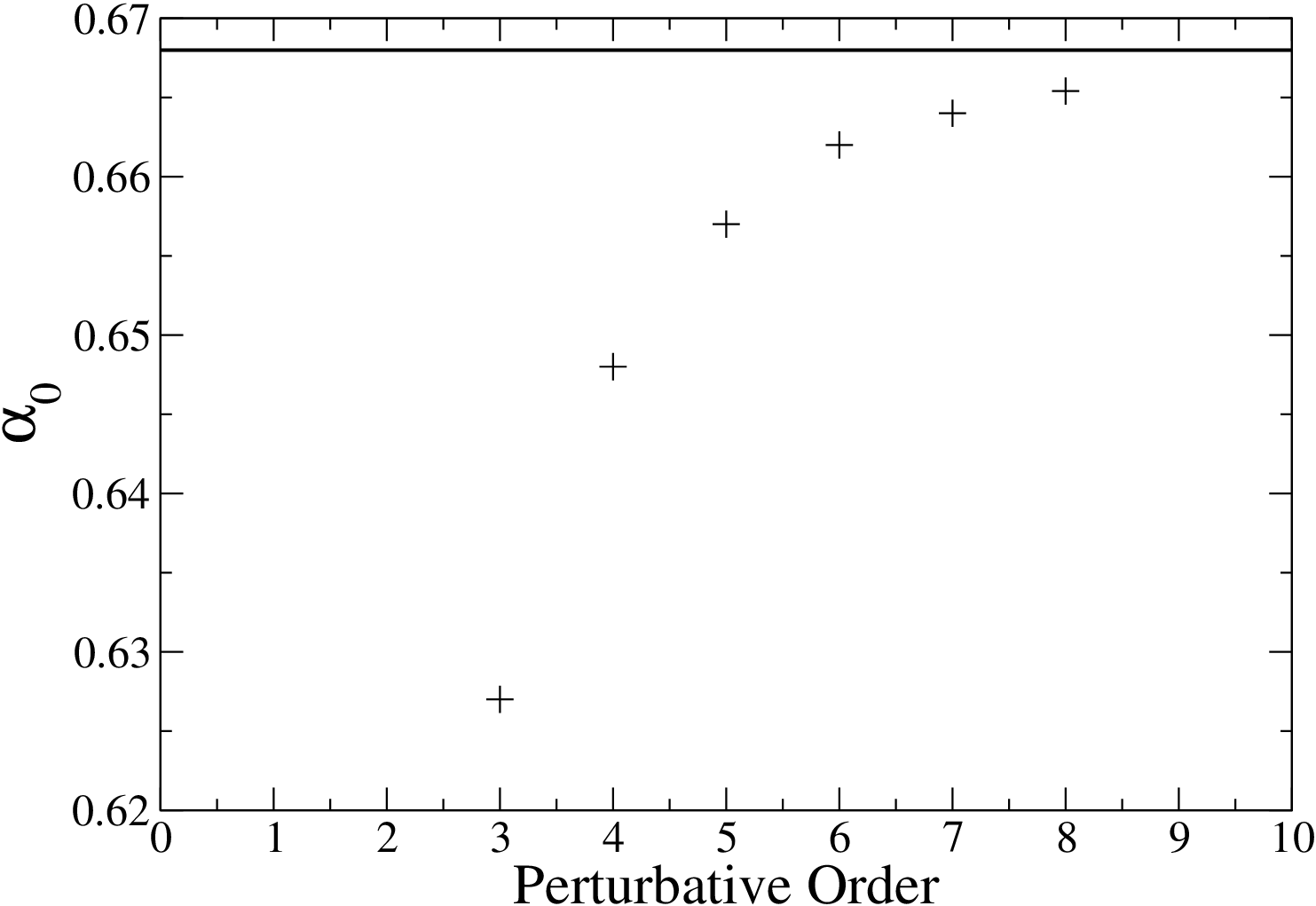}
\caption{The coefficient $\alpha_0$ in the asymptotic expansion of the ground
  state energy (see text) at different perturbative orders, up to order $8$. 
  The solid line is the exact result \protect\cite{Janke:1995wt}.}
\label{fig:alpha}
\end{figure}

The extension of the method to include higher orders is straightforward and the
details will be presented elsewhere \cite{AAD03b}. 
Here we just present the expression for the energy up to third order
\beq
  E_{0}^{(3)} &=& \frac{\hbar}{32 m^2 {\to^5}} 
    \left[6 \mu \hbar \to \left( - {\omega}^2 + 2 {\to}^2 \right) + m^2 \left(
    \omega^6 \right. \right. \nonumber \\
  &&\quad - \left. \left. 5 \omega^4 \to^2 + 15 \omega^2 \to^4 + 5 \to^6
    \right) \right].
\label{eq:ord3en}
\eeq
The optimal value of $\Omega$ is then obtained by using the PMS. 
The expression for $\Omega$ in the general case is lengthy (see \cite{AAD03b})
and we only write here its asymptotic value in the limit of large $\mu$ (from
now on we assume  $\hbar = m = \omega = 1$):
\beq
  \Omega = 2 \left( \frac{3 \mu}{5}\right)^{1/3} + O\left[ {\mu^{-1/3}} \right].\eeq
We can then substitute this expression in Eq.~(\ref{eq:ord3en}) and extract the
asymptotic behavior of the energy of the ground state to third order:
\beq
  E_{0}^{(3)} &=& \frac{3}{16} \left(\frac{75}{2} \right)^{1/3} 
    \left(\frac{\mu}{4} \right)^{1/3} + O\left[ {\mu^{-1/3}} \right] .
\label{engs3}
\eeq
The asymptotic behavior of the ground state energy of the AHO has been studied
in \cite{Janke:1995wt}, by using the exact Rayleigh-Schr\"odinger perturbation
coefficients as an input. We use this calculation as a reference to compare to 
our results. Following the notation set in \cite{Janke:1995wt} we write the
energy as 
\beq
  E_{0} &=& \alpha_0 \left(\frac{\mu}{4} \right)^{1/3} + O\left[ {\mu^{-1/3}} 
    \right] ,
\eeq
and using this formula in Eq.~(\ref{engs3}) we obtain the coefficient
$\alpha_0$ to third order in our expansion to be 
\beq
  \alpha_0^{(3)} = \frac{3}{16} \left(\frac{75}{2} \right)^{1/3} \approx 0.627,
\eeq
which falls within $6~\%$ of the true value. We have also computed $\alpha_0$ up
to eighth order, where the agreement is within $0.4~\%$. 
The results are shown if Fig.~\ref{fig:alpha}.

\begin{figure}
\includegraphics[width=8.5cm]{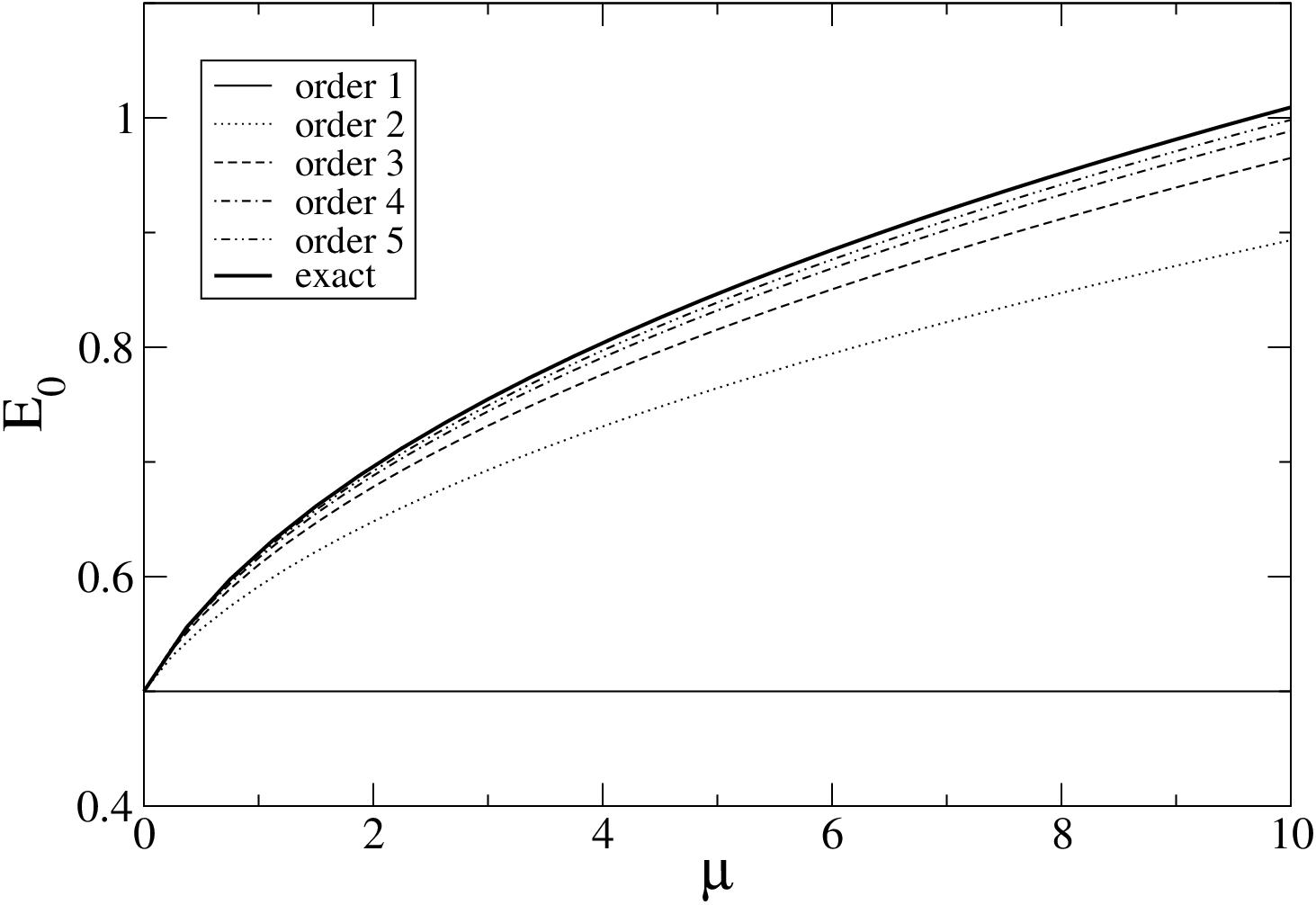}
\caption{Ground state energy of the AHO as a function of $\mu$ to different
  perturbative orders, assuming $\hbar = m = \omega = 1$. }
\label{fig:GSmu}
\end{figure}

In Fig.~\ref{fig:GSmu} we plot the energy of the AHO as a function of $\mu$.
The solid bold curve is the exact result obtained by numerically solving the
Schr\"odinger equation for the AHO. The other curves are the energies computed
at different perturbative orders, up to fifth order (from bottom to top). 
We notice from the plot that the perturbative result is always below the true
result; this is to be contrasted with a variational calculation, where the
exact result would be approached from above. We do not know whether this is a
general property.   

\begin{figure}[t]
\includegraphics[width=8.5cm]{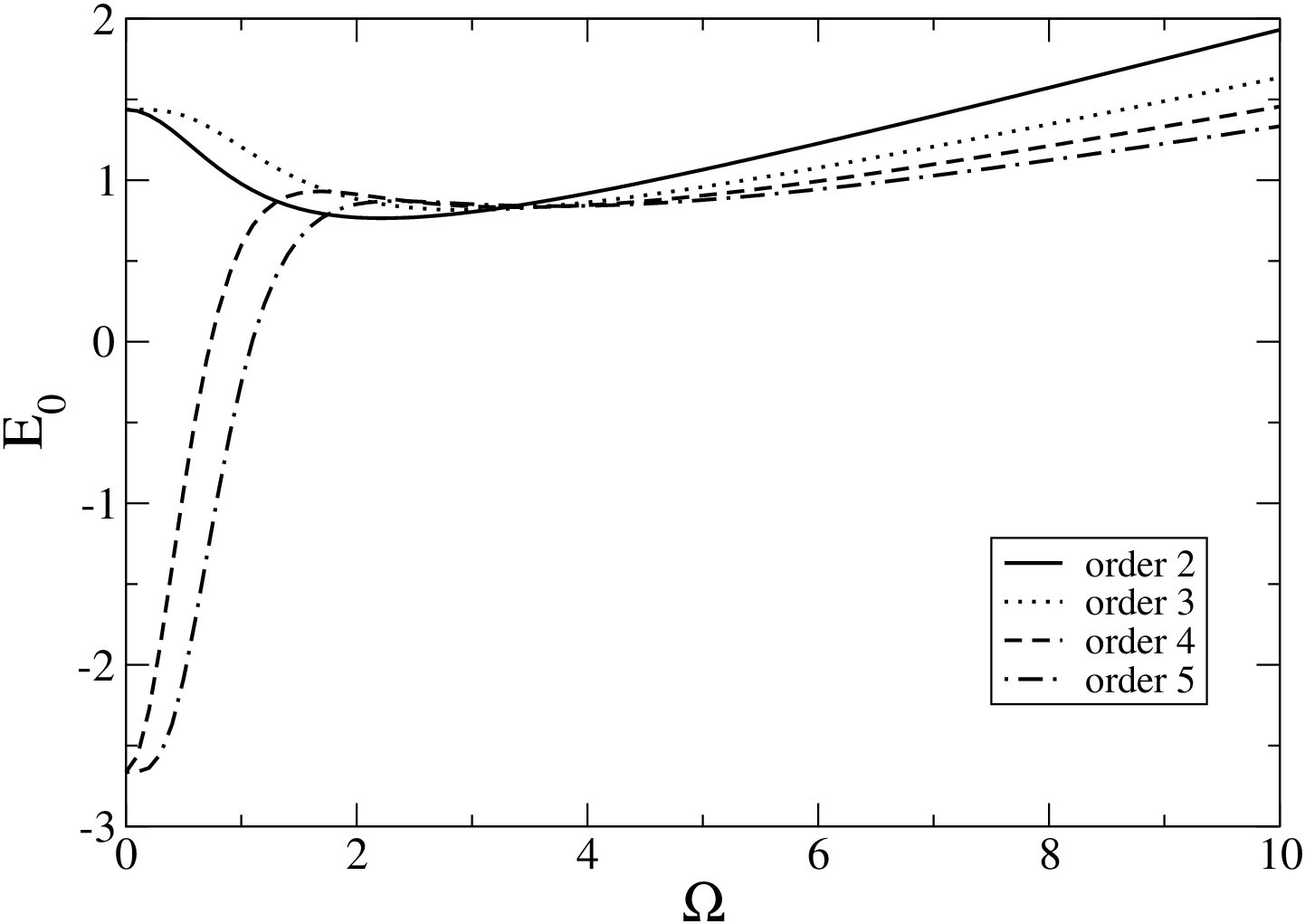}
\caption{Energy of the AHO as a function of $\Omega$, at different perturbative
  orders ($\hbar = m = \omega = 1$) and keeping $\mu$ fixed ($\mu = 5$). }
\label{fig:GSomega}
\end{figure}

\begin{figure}
\includegraphics[width=8.5cm]{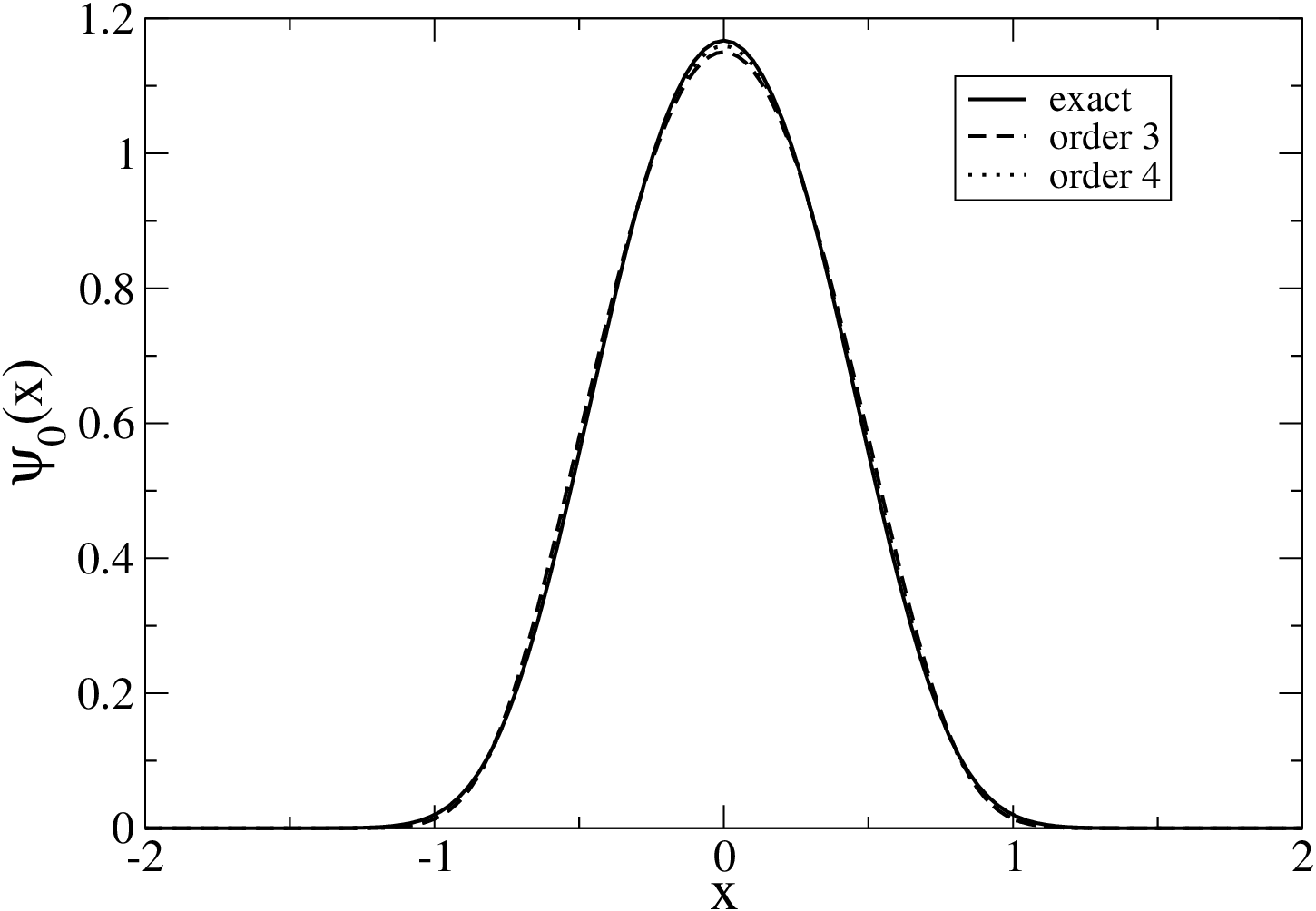}
\bigskip
\caption{Wave function for the ground state of the anharmonic oscillator,
  assuming $m = \hbar = \omega = 1$ and $\mu = 200$. 
  The solid line is the exact result whereas the dashed (dotted) line is the
  result obtained with our method to third (fourth) order.} 
\label{fig:psi}
\end{figure}

In Fig.~\ref{fig:GSomega} we show the energy of the AHO as a function of the
parameter $\Omega$, for $\mu = 5$. 
The different curves correspond to different orders in perturbation theory
(from second up to fifth order). 
We observe that the energy at different orders in the perturbative expansion 
develops local minima and maxima. By applying the PMS we select the extremum in
which the energy is flatter, i.~e. where the dependence upon $\Omega$ is
smaller. 

It should be remarked that the method works equally well for the wave function.
Indeed, in Fig.~\ref{fig:psi} we plot the wave function of the ground state of
the AHO. The solid line is the exact (numerical) result, whereas the dashed and
dotted lines refer to the wave function obtained by applying our method to
third and fourth order respectively. We see that it works extremely well, even
for large values of $\mu$ ($\mu = 200$).  
Although we have not imposed that the wave function be flat at the origin, we
observe that the method naturally provides this result.

Finally, we would like to stress that, although we have displayed results up to
a few orders in the perturbative expansion, it is very easy to push the
calculation to any order, since the method only requires the solution of
algebraic equations order by order.

We are currently working on the application of our method to the calculation of
the excited states of the AHO, to more general anharmonic potentials and to the double-well potential.

\begin{acknowledgments}
P.A. and A.A. acknowledge support for this work to the ``Fondo Alvarez-Buylla''
of Colima University. P.A. also acknowledges Conacyt grant no. C01-40633/A-1.
\end{acknowledgments}

\end{document}